\documentclass{PoS}
\usepackage[normalem]{ulem}

\title{Isospin symmetry breaking effects in the pion and nucleon masses}

\ShortTitle{Isospin symmetry breaking effects in the pion and nucleon masses}

\author{\speaker{Ran~Zhou}\thanks{We thank Riken, Brookhaven National
 Laboratory and the U.S. Department of Energy for providing the 
facilities and hospitality where this work was done.}\\
  Physics Department, University of Connecticut, Storrs, CT 06269-3046
 , USA\\
        E-mail: \email{zhouran@phys.uconn.edu}}

\author{Thomas~Blum\\
        Physics Department, University of Connecticut, Storrs, CT
        06269-3046, USA\\
        RIKEN BNL Research Center, Brookhaven National Laboratory,
        Upton, New York 11973, USA\\
        E-mail: \email{tblum@phys.uconn.edu}}

\author{Takumi~Doi\\
        Department of Physics and Astronomy, University of Kentucky,
        Lexington, KY 40506, USA\\
        E-mail: \email{doi@ribf.riken.jp}}

\author{Masashi Hayakawa\\ 
        Department of Physics, Nagoya University,
        Nagoya 464-8602, Japan\\
        E-mail: \email{hayakawa@eken.phys.nagoya-u.ac.jp}}

\author{Taku~Izubuchi\\
        Institute for Theoretical Physics, Kanazawa University,
        Kanazawa 920-1192, Japan\\
        RIKEN BNL Research Center, Brookhaven National Laboratory,
        Upton, New York 11973, USA\\
        E-mail: \email{izubuchi@quark.phy.bnl.gov}}

\author{Norikazu~Yamada\\
        High Energy Accelerator Research Organization(KEK), Tsukuba,
        Ibaraki 305-0801, Japan \\
        The Graduate University for Advanced Studies, Tsukuba,
        Ibaraki 305-0801, Japan \\
        E-mail: \email{norikazu.yamada@kek.jp}}

\abstract{We investigate the electromagnetic mass splittings in 
{the pseudoscalar meson and nucleon systems} by combining 2+1 
flavor domain wall fermion 
gauge configurations, generated by the RBC and UKQCD collaborations, 
and quenched, non-compact, lattice QED configurations. We 
analyze finite volume effects by using {$16^3\times 32$ 
and $24^3\times 64$ lattices}.}

\FullConference{The XXVI International Symposium on Lattice Field Theory\\
		 July 14-19 2008\\
		 Williamsburg, Virginia, USA}

\begin{document}

\section{Introduction}

The mass splitting in the meson and baryon system is an interesting 
topic in hadron spectroscopy. It is related with the quark masses which
are the fundamental parameters of the Standard Model. The mass
difference of the proton and neutron is also crucial to the
phenomenological model of nuclei, because it plays an important role 
in the neutron $\beta$ decay process. These mass differences 
have already been measured by experimentalists to  good accuracy.
So they can be used as input to help theorists to determine
the parameters in their phenomenological models or check the
validity of the theory.

The mass of the hadron is determined by both QED and QCD
dynamics. For the QCD interaction, since the coupling constant 
is too large at the low energy regime, the perturbation theory 
is not applicable. People have developed all kinds of 
effective theories of QCD from the 1950s. They can be used to describe the 
hadron mass spectrum. For instance, the mass of the 
pseudoscalar meson, especially for the pion, is much smaller 
than the mass of the baryon. It can be explained by the
spontaneous breakdown of chiral symmetry. When QED is included, the meson
masses are split, but
$m^2_{\pi^+}-m^2_{\pi^0}\simeq m^2_{K^+}-m^2_{K^0}$, which 
is explained by Dashen's 
theorem~\cite{Dashen:1969}. Partially
quenched chiral perturbation theory(PQ$\chi$PT) gives a
more general description of QCD phenomena:  
valence and sea quark masses are varied independently. It is
important since the cost to generate the dynamical QCD configurations 
for a large lattice is still pretty high today. 

Recent developments in PQ$\chi$PT~\cite{Bijnens:2007} 
allows us to incorporate the QED interaction too, so we can 
investigate the mass splitting which is related to the QED interaction.
Since the QED interaction is much  weaker than the QCD interaction, 
it will only provide a small shift in the hadron mass from 
its QCD value. But this shift is important to the splitting in the
pion and kaon multiplet.

Because the lattice calculation can give the hadron mass spectrum from
first principles, it is used to test whether the QCD+QED
interaction can actually reproduce the hadron spectrum. The pioneering 
work on this topic has been done in~\cite{Duncan:1996} with the
quenched QCD and non-compact QED configuration. The development of 
Lattice QCD enables us to investigate the hadron spectrum 
with 2+1 flavor dynamical QCD configurations, which means we can include
the effect of the degenerate up and down quarks and also the heavier
strange quark in the sea quark sector. So our simulation
result can be used to fit the PQ$\chi$PT formula. We can extract the
unknown low energy constants(LEC's) from the fitting. After we know the LEC's,
we can use the pseudoscalar meson mass as input to determine the 
physical value of the $u$, $d$, and $s$ quark masses. {{The 
MILC group also reports their analysis of the mass splitting} using
staggered fermion ensembles in Basak's Talk.}

The lattice simulation is also used for the mass splitting in the 
nucleon system~\cite{Savage:2007}. We present preliminary
results for the mass difference between the proton and neutron due to QED.

\section{Theoretical Background}
In chiral perturbation theory of QCD, the mass square of the pseudoscalar
meson is determined by the spontaneous breakdown of chiral symmetry from 
$SU(3)_L \otimes SU(3)_R \otimes U(1)_V$ to $SU(3)_V \otimes U(1)_V$. 
The Goldstone boson generated by this mechanism is the 
pseudoscalar meson octet. All of them are massless in the chiral limit.
If the QED interaction is also considered, the symmetry group
will be broken further to $SU(2)_V \otimes U(1)_{em}\otimes U(1)_V$. 
The only Goldstone bosons left are the  neutral pion and kaons (the neutral pion
can be considered a Goldstone boson if ${\cal O}(\alpha^2)$ terms are neglected\cite{Blum:2fEM}). 

Since we calculate with several valence quark mass and charge 
combinations, we use PQ$\chi$PT. 
Suppose we have a ``kaon" which is composed by {$u$ and $s$ 
quarks}. Following the 
notation of \cite{Bijnens:2007},
index 1 denotes the u quark and 3 the s quark. The square
of the meson mass is\cite{Bijnens:2007}

\begin{eqnarray} 
m^2 =\chi_{13}+\frac{2Ce^2}{F_0^2}\left(q_1-q_3\right)^2+
\frac{\delta^{(4)}}{F^2_0}
\label{equ:m21}
\end{eqnarray} 
\begin{eqnarray} 
\frac{\delta^{(4)}}{F^2_0} &=&
 \left[(48 L_6^r - 24 L_4^r) \bar \chi_1 \chi_{13}
+  (16 L_8^r - 8 L_5^r) \chi_{13}^2 
-\frac{1}{3} \bar A(\chi_m) R^m_{n13} \chi_{13}
-\frac{1}{3} \bar A(\chi_p) R^p_{q\pi\eta} \chi_{13}\right]/{F^2_0}
\nonumber \\ &&
+  e^2 \bar A(\chi_{13}) q_{13}^2 +4  e^2 \bar B(\chi_\gamma,\chi_{13},\chi_{13}) q_{13}^2
\chi_{13}-4 e^2 \bar B_1(\chi_\gamma,\chi_{13},\chi_{13}) q_{13}^2 \chi_{13}
\nonumber \\ &&
-{C}e^2[-48 L_4^r q_{13}^2 \bar \chi_1
-16 L_5^r q_{13}^2 \chi_{13}
+2 \bar A(\chi_{1s}) q_{1s} q_{13}
-2 \bar A(\chi_{3s}) q_{3s} q_{13}]/F_0^4\nonumber \\ &&
- {Y_1} 4 e^2 \bar{Q_2}\chi_{13}+ {Y_2} 4 e^2 q_p^2\chi_p+
{Y_3} 4 e^2 q_{13}^2 \chi_{13}
- {Y_4} 4 e^2  q_1 q_3\chi_{13}
+ {Y_5} 12 e^2 q_{13}^2 \bar \chi_1\nonumber \\ &&
\label{equ:m22}
\end{eqnarray}

$Y_i$ are independent linear combinations of the LEC's $K^{Er}$ as defined in Eq. (48) of
\cite{Bijnens:2007}, and  $\chi_i$ is the LO meson
mass,
{$\chi_i=2 B_0 m_i,~\chi_{ij}=(\chi_i+\chi_j)/2$, $q_{13}
=q_1-q_3=q_u-q_s$.}
 The form of the functions $\bar{A}, \bar{B}, \bar{B}_1, R$, 
and others can be found in~\cite{Bijnens:2007}.

\begin{table}[h]
 \centering
 \begin{tabular}{|c|c|c|}
  \hline  
   & QCD LEC's& QED LEC's  \\
  \hline
 LO & $B_0, F_0$ & $C$ \\
  \hline
 NLO & $L^r_4, L^r_5, L^r_6, L^r_8$ & $C, Y_1, Y_2, Y_3, Y_4, Y_5$ \\  
  \hline
 \end{tabular}
 \caption{LEC's in PQ$\chi$PT}
 \label{tab:LEC's}
\end{table}
Table \ref{tab:LEC's} shows all of the LEC's which are necessary to
determine the pseudoscalar meson mass at NLO. The QCD LEC's have been computed
in\cite{RBC:24}, so the current simulation is used to extract 
the QED LEC's. Note that the definition of the decay constant used in~\cite{Bijnens:2007} (followed here) is $\sqrt{2}$ smaller than the one in~\cite{RBC:24}. 
For the splitting in baryon system, we focus on the mass differences
between the proton and neutron.
Here, we study only the degenerate case,  $m_u=m_d$. Then the leading contribution
to the mass difference
is proportional to $\alpha$. 

\section{Lattice Simulation}

Following\cite{Duncan:1996}, the lattice calculation employs combined QCD+QED gauge configurations. 
For the QCD configurations, we use the $N_f=2+1$ flavor QCD configurations
generated with the domain wall fermion and Iwasaki gauge actions
 by the RBC and UKQCD collaborations~\cite{RBC:16.1, RBC:16.2,
  RBC:24}. 
{For the $16^3\times 32$ lattice}, we use light quark mass
0.01, 0.02 and 0.03 ensembles. {For $24^3\times 64$ lattice}, 
we use the 0.005, 0.01, 0.02 and 0.03 ensembles. The strange quark
mass is fixed to 0.04 in all cases. {Table \ref{Tab:ensembles}
gives the details of the lattice ensembles used in this
work.}

The QED configurations are generated in a quenched, non-compact 
form\cite{Blum:2fEM,Hayakawa:2008an}.  Here we
employ the Feynman gauge instead of the Coulomb gauge
which was used previously\cite{Blum:2fEM}. 
One advantage of the {non-compact QED formalism is that 
the $U(1)$ gauge potential $A_\mu$ can be chosen }randomly with the 
correct distribution
in momentum space, then Fourier transformed to coordinate space, 
so there are no autocorrelations in the ensemble. Another
advantage is that there is no (lattice artifact) photon self-interaction in the action. 
{To couple $A_\mu$ to the fermions}, the non-compact field is exponentiated in the usual way, 
so the combined gauge field appears as an $SU(3)$ matrix times a phase, 
$U^{QCD}_\mu(x){\times}U^{QED}_\mu(x)$.

Since the QED interaction does not have confinement, it is possible 
that the finite volume can induce a 
significant systematic error. So, we do our simulation on both 
$16^3$ and $24^3$ lattice configurations with the same lattice spacing.
Then we can investigate the finite volume effect in the mass  spectrum.

\begin{table}
 \centering
 \begin{tabular}{ccccccc}
  \hline  
  lat & $m_{\rm sea}$ & $m_{\rm val}$ & Trajectories & $\Delta$ & $N_{meas}$ & $t_{src}$\\
  \hline
  $16^3$ & 0.01 & 0.01-0.03 & 500-4000 & 20  & 176 & 0 \\
  $16^3$ & 0.02 & 0.01-0.03 & 500-4000 & 20  & 176 & 0 \\
  $16^3$ & 0.02 & 0.01-0.03 & 500-4000 & 20  & 176 & 0 \\
  \hline
  $24^3$ & 0.005 & 0.005-0.03 & 900-8060 & 40  & 180 & 0 \\
  $24^3$ & 0.01 & 0.01-0.03 & 1460-5040 & 20 &  180 & 0 \\
  $24^3$ & 0.02 & 0.02 & 1800-3580 & 20 & 180 & 0,32 \\
  $24^3$ & 0.03 & 0.03 & 1800-3580 & 20 & 180 & 0,32 \\
  \hline
 \end{tabular}
 \caption{QCD gauge configuration ensembles generated by
   the RBC and UKQCD collaborations~\cite{RBC:16.1, RBC:16.2, RBC:24}. 
   $\Delta$ is the separation of the 
   trajectory number in molecular dynamics time units. 
   $\beta_{\tiny\rm QCD}=2.13$. The inverse lattice scale is 
   $a^{-1}=1.729(28)$GeV for both of $16^3$ and $24^3$ lattice.}
 \label{Tab:ensembles}
\end{table}

\section{Numerical Results}

We can switch on and off the QED interaction by setting the charge
$e\neq 0$ and $e=0$. We first extract the masses from wall source, 
point sink, correlation functions. We define the mass difference, 
$\delta m^2=m^2({e\neq 0})-m^2({e=0})$, and then fit $\delta m^2$ with 
the PQ$\chi$PT formula, using the standard jackknife method to
estimate the statistical error. {$\delta m^2$ is averaged over  
$+e$ and $-e$ ensembles to decrease the noise
in the signal~\cite{Blum:2fEM}.}
We take the QCD LEC's from
the $SU(3)$ PQ$\chi$PT fit in~\cite{RBC:24} since we use the same ensembles.
We also use these for the $16^3$ case because these two
ensembles have the same lattice scale. Then we fit our data
to obtain all of the QED LEC's, except $Y_1$ because the sea quarks are not
coupled to photons. Thus, $Y_1=0$ in this work.

PQ$\chi$PT is based on chiral symmetry,
which is not exact in the domain wall fermion formalism
with finite $L_s$. To lowest order, the violation of chiral symmetry can 
be counted as a small shift in the input quark mass, so we 
substitute $m_i+m_{res}$ for all quark masses, where
the residual mass $m_{res}$ is determined from the method
in reference ~\cite{Blum:2fEM}. Also following~\cite{Blum:2fEM},
we add additional LEC's, $C_1$ and $C_2$, proportional to
$(q_1-q_3)^2$ and $(q_1+q_3)^2$, respectively,
to account for chiral symmetry breaking due to finite $L_s$ at order ${\cal O}(\alpha)$. 
These LEC's vanish {when $L_s\to\infty$. $C_1$ mixes with the
  LEC $C$ in Eq. (\ref{equ:m22})}, 
so it must be subtracted to obtain the physical LEC. $C_1$ is measured from $m_{res}$ with $e\neq0$ as
described in \cite{Blum:2fEM}. For the preliminary results presented here, this subtraction
has not been performed.

\begin{figure}[ht]
\begin{minipage}[b]{0.4\linewidth}
\hskip -1cm
\includegraphics[scale=0.8]{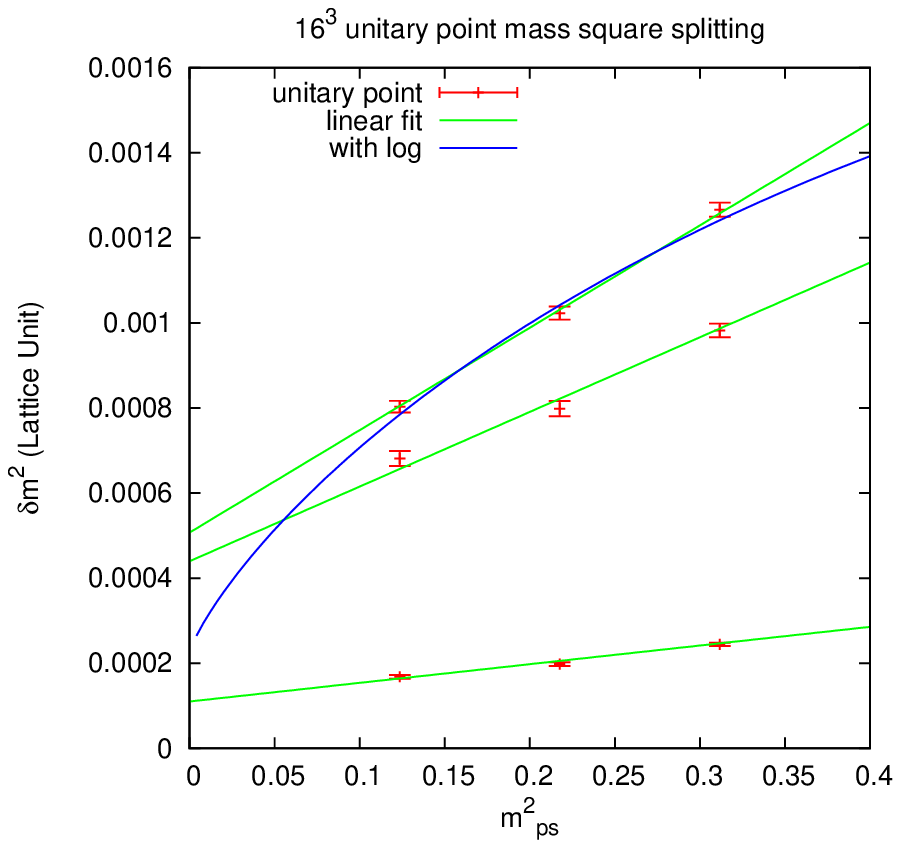}
\end{minipage}
\hspace{.5cm}
\begin{minipage}[b]{0.5\linewidth}
\includegraphics[scale=0.8]{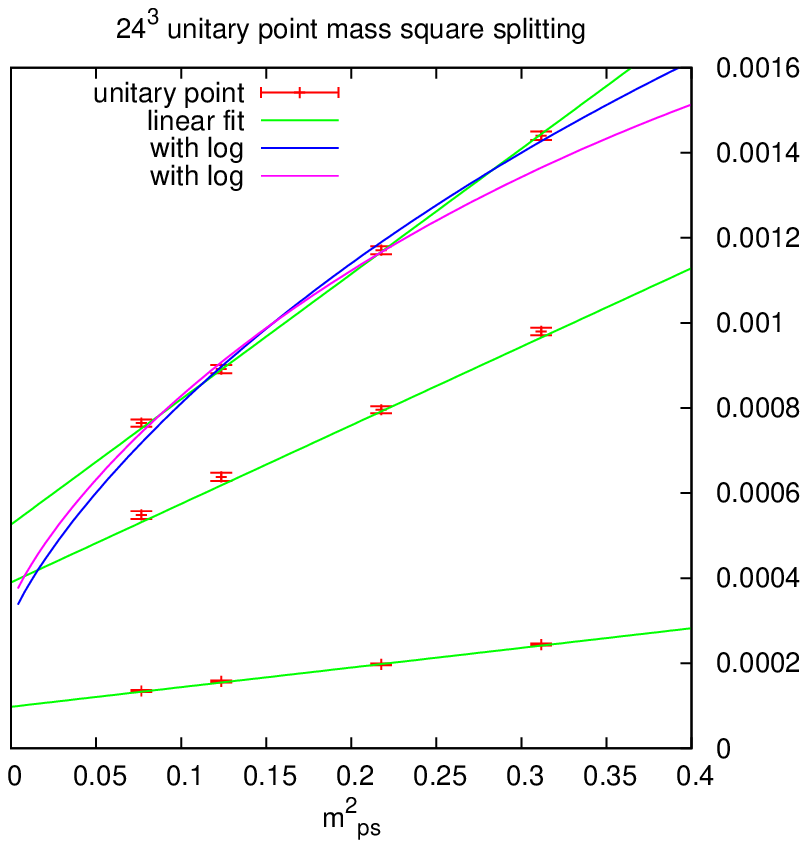}
\end{minipage}
\caption{$16^3$ (left panel) and $24^3$ (right panel) lattice data
 points and fit results for the meson mass splitting. Lines
 correspond to fits with and without chiral logs. {The data 
 points in the plot correspond to $u\bar d, u\bar u$ and $d\bar d$
 mesons, respectively, from top to bottom.}}
\label{Fig:1624}
\end{figure}

By fitting the results for $\delta m^2$, we can get all of the QED LEC's, including the pure lattice artifact $C_2$. The fit results are shown in Figure~\ref{Fig:1624} along with the unitary data points (though all of the partially quenched data was used in the fit) and tabulated in Table~\ref{tab:LEC}. We perform fits with and without chiral logs. $\chi^2$/dof is adequate in either case, except for $24^3$ when all the quark mass data is fitted. If the heaviest sea quark points are omitted, $\chi^2$ becomes reasonable.  The main effect of the logs is to significantly reduce the value of the charged meson 
splitting in the chiral limit. Note, the log terms vanish for the neutral mesons, and the neutral splittings should also
vanish in the chiral limit (see \cite{Blum:2fEM} for a detailed
discussion). {The non-vanishing chiral limit value in the plot is 
 due of the $C_2$ term, or the explicit chiral symmetry breaking
 induced by finite $L_s$.}

\begin{table}[h]
 \centering
 \begin{tabular}{|c|c|c|c|c|c|c|}
  \hline  
  \ & $16^3$ & $24^3$ & $16^3$ & $24^3$ & $24^3$ & Ref.~\cite{Bijnens:2007}\\
  \hline
  with log & No & No & Yes & Yes & Yes & Yes \\
  \hline
  fit range & 0.01-0.03 & 0.005-0.03 & 0.01-0.03 & 0.005-0.03 &
  0.005-0.02 & N/A\\
  \hline
  $C\times 10^6$ & 3.48(40) & 3.02(14) & 0.85(16) & 0.769(64) &
  0.96(10) & 2.71\\
  \hline
  $Y_2\times 10^2$ & 1.47(3) & 1.51(1) & 1.56(3) & 1.54(1) & 1.51(2) &
  0.53 \\
  \hline
  $Y_3\times 10^3$ & 7.64(72) & 7.61(39) & -1.08(45) & -1.17(35) &
  -2.79(43) & 2.65\\
  \hline
  $Y_4\times 10^3$ & 8.00(78) & 7.55(50) & 9.71(92) & 8.22(51) &
  8.85(80) & 5.7\\
  \hline
  $Y_5\times 10^3$ & 1.48(23) & 2.43(12) & 1.05(24) & 2.23(12) &
  2.35(19) & 0\\
  \hline
  $C_2\times 10^3$ & 2.69(10) & 2.39(5) & 2.69(10) & 2.39(5) & 2.48(6)
  & N/A\\
  \hline
  $\chi^2/dof$ & 0.7(6) & 0.6(2) & 1.0(6) & 2.8(5) & 1.2(6) & N/A\\
  \hline
 \end{tabular}
 \caption{The QED LEC's from fits of $\delta m^2$. The pure QCD LEC's were taken from the SU(3) fit in~\cite{RBC:24}. The last column gives phenomenological results~\cite{Bijnens:2007}.}
 \label{tab:LEC}
\end{table}

From Table~\ref{tab:LEC}  we can see that the LEC's change slightly from 
$16^3$ to $24^3$, but  $C$ and $Y_3$ change significantly when the 
the log terms are included. The effect on the other LEC's is not so
drastic. Also shown in Table~\ref{tab:LEC} are the values from 
phenomenology~\cite{Bijnens:2007}, {which are roughly consistent with 
ours within an order of magnitude}, though we emphasize that our values 
do not include systematic 
corrections due to finite volume, explicit chiral symmetry breaking, 
and non-zero lattice spacing, to name just three, and so should be 
taken as preliminary.

\begin{figure}[ht]
\begin{minipage}[b]{8cm}
\hskip -1.5cm
\includegraphics[scale=0.8]{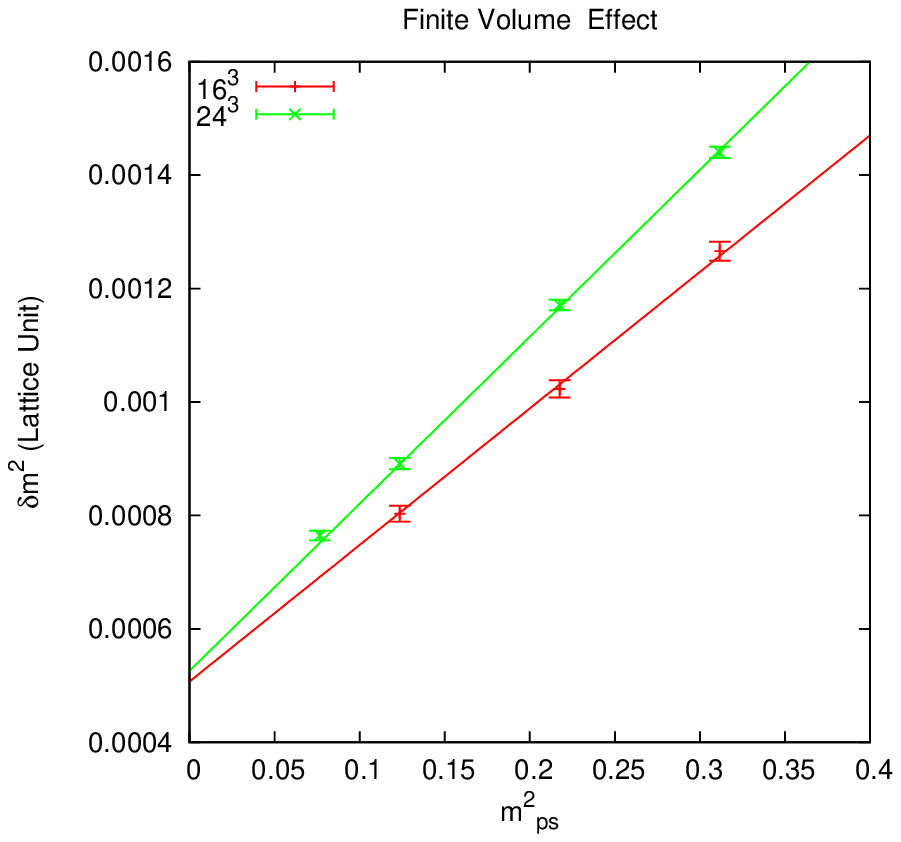}
\caption{Finite volume effect on the $u\bar d$ meson.}
\label{Fig:comp}
\end{minipage}
\begin{minipage}[b]{8cm}
\hskip -2cm
\includegraphics[scale=0.8]{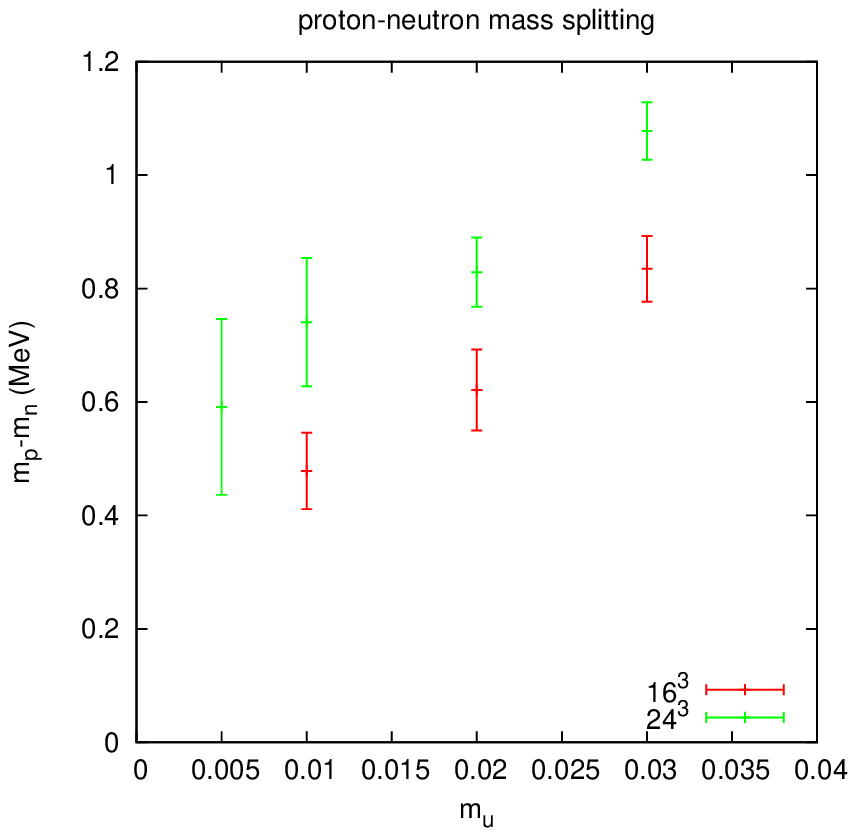}
\caption{The proton-neutron mass difference.}
\label{Fig:pn}
\end{minipage}
\end{figure}

Figure \ref{Fig:comp} shows the comparison of the charged meson splitting from 
$16^3$ and $24^3$ lattices, and the relative change 
is given in Table~\ref{Tab:finitev}. The difference actually decreases at
the smallest quark mass, 0.01, which is why the finite volume effect
shown in Figure~\ref{Fig:comp} appears to go away in the chiral limit
(linear fit). Interestingly, from Figure~\ref{Fig:1624}, one can see that the neutral splitting is mostly insensitive to the finite volume, at least for the volumes studied here.
We are still investigating these effects, and plan a detailed  comparison with finite volume chiral perturbation theory\cite{Hayakawa:2008an}.

\begin{table}[h]
 \centering
 \begin{tabular}{|c|c|}
  \hline  
  $m_{val}$ & $2\frac{\delta m^2(24^3)-\delta m^2(16^3)}{\delta m^2(24^3)+\delta m^2(16^3)}$  \\
  \hline
  0.01 & 0.1043(9) \\
  \hline
  0.02 & 0.1349(9) \\
  \hline
  0.03 & 0.1286(9)\\
  \hline
 \end{tabular}
 \caption{Finite volume effect on the $u\bar d$ meson.}
 \label{Tab:finitev}
\end{table}

Figure~\ref{Fig:pn} is the result of the mass difference between the proton
and neutron when the QED interaction is included. If there is no QED
interaction and $m_u=m_d$, then $m_n=m_p$, which is the result of isospin symmetry. Figure~\ref{Fig:pn} shows the
mass difference for the unitary points.
We can see the proton is heavier than the 
neutron and the mass difference
decreases with quark mass. 
The
$24^3$ result is larger than the $16^3$ result, once again signaling finite volume corrections. 
The baryon mass difference is noisier than in the pseudoscalar case. 
We plan to increase the statistics for 
the baryon mass spectrum to improve the precision on
this result, and to compute the mass splitting coming from non-degenerate $u$ and $d$ quark masses, which is expected to switch the sign of the mass difference, in accord with Nature.

\section{Summary and outlook}
Using 2+1 flavor QCD and non-compact QED,
we have studied the electromagnetic mass splitting in hadrons. 
We have done our
simulation on different sea quark mass ensembles and have used 
both $16^3$ and $24^3$ lattices to investigate the finite volume
effect on the mass splitting. 

Preliminary fits appear to be consistent with 
NLO PQ$\chi$PT, including photons, for the parameters in our study, except at the heaviest quark mass, 0.03, which is close to the physical strange quark mass.
A more complete study of systematic errors, including finite volume and explicit chiral symmetry breaking effects, is in progress.
Next we will use the LEC's obtained from our analysis to determine 
the physical, non-degenerate, $u$, $d$, and $s$ quark masses.

\section*{Acknowlegements}

We thank the US Department of Energy and RIKEN for the support necessary to carry out this research. RZ  and TB were supported by US DOE grant  DE-FG02-92ER40716. Computations were carried out on the QCDOC supercomputers at the RIKEN
BNL Research Center, BNL, and Columbia University.

\end{document}